\documentclass[aps,prb,superscriptaddress,showpacs,citeautoscript,twocolumn,longbibliography]{revtex4-1}

\usepackage{graphicx}
\usepackage{amsmath}
\usepackage{amssymb}
\usepackage{color}
\usepackage{hyperref}
\usepackage{ulem}         

\newcommand{\bea}{\begin{eqnarray*}}
\newcommand{\eea}{\end{eqnarray*}}
\newcommand{\bne}{\begin{equation*}}
\newcommand{\ede}{\end{equation*}}

\newcommand{\bnen}{\begin{equation}}
\newcommand{\eden}{\end{equation}}
\newcommand{\bean}{\begin{eqnarray}}
\newcommand{\eean}{\end{eqnarray}}
\newcommand{\bnsn}{\begin{subequations}}
\newcommand{\edsn}{\end{subequations}}

\newcommand{\bna}{\begin{array}}
\newcommand{\eda}{\end{array}}
\newcommand{\bnm}{\begin{enumerate}}
\newcommand{\edm}{\end{enumerate}}

\newcommand{\ket}[1]{| #1 \rangle}
\newcommand{\bra}[1]{\langle #1 |}

\definecolor{gray}{rgb}{.6,.6,.6}

\begin{document}

\title{From Cooper pair splitting to the non-local spectroscopy of a Shiba state}

\author{Zolt\'an~Scher\"ubl}
%\email{scherubl.zoltan@gmail.com}
\affiliation{Department of Physics, Budapest University of Technology and Economics and MTA-BME "Momentum" Nanoelectronics Research Group, H-1111 Budapest, Budafoki \'ut 8., Hungary}
\affiliation{Univ. Grenoble Alpes, CEA, Grenoble INP, IRIG, PHELIQS, 38000 Grenoble, France}
\author{Gerg\H{o}~F\"ul\"op}
\affiliation{Department of Physics, Budapest University of Technology and Economics and MTA-BME "Momentum" Nanoelectronics Research Group, H-1111 Budapest, Budafoki \'ut 8., Hungary}
\author{J\"org Gramich}
\affiliation{Department of Physics, University of Basel, Klingelbergstrasse 82, CH-4056 Basel, Switzerland}
\author{Andr\'as~P\'alyi}
\affiliation{Department of Theoretical Physics and MTA-BME Exotic Quantum Phases Research Group, Budapest University of Technology and Economics, H-1111 Budapest, Hungary}
\author{Christian~Sch\"onenberger}
\affiliation{Department of Physics, University of Basel, Klingelbergstrasse 82, CH-4056 Basel, Switzerland}
\author{Jesper~Nyg\r{a}rd}
\affiliation{Center for Quantum Devices and Nano-Science Center, Niels Bohr Institute, University of Copenhagen, Universitetsparken 5, DK-2100 Copenhagen, Denmark}
\author{Szabolcs~Csonka}
%\email{csonka@mono.eik.bme.hu}
\affiliation{Department of Physics, Budapest University of Technology and Economics and MTA-BME "Momentum" Nanoelectronics Research Group, H-1111 Budapest, Budafoki \'ut 8., Hungary}

\begin{abstract}
Cooper pair splitting (CPS) is a way to create spatially separated, entangled electron pairs. To this day, CPS is often identified in experiments as a spatial current correlation. However, such correlations can arise even in the absence of CPS, when a quantum dot is strongly coupled to the superconductor, and a subgap Shiba state is formed. Here, we present a detailed experimental characterization of those spatial current correlations, as the tunnel barrier strength between the quantum dot and the neighboring normal electrode is tuned. The correlation of the non-local signal and the barrier strength reveals a competition between CPS and the non-local probing of the Shiba state. We describe our experiment with a simple transport model, and obtain the tunnel couplings of our device by fitting the model's prediction to the measured conductance correlation curve. Furthermore, we use our theory to extract the contribution of CPS to the non-local signal. 
\end{abstract}

\date{\today}
\maketitle

\section{Introduction}

\begin{figure}[!htbp]
	\begin{center}
	\includegraphics[width=\columnwidth]{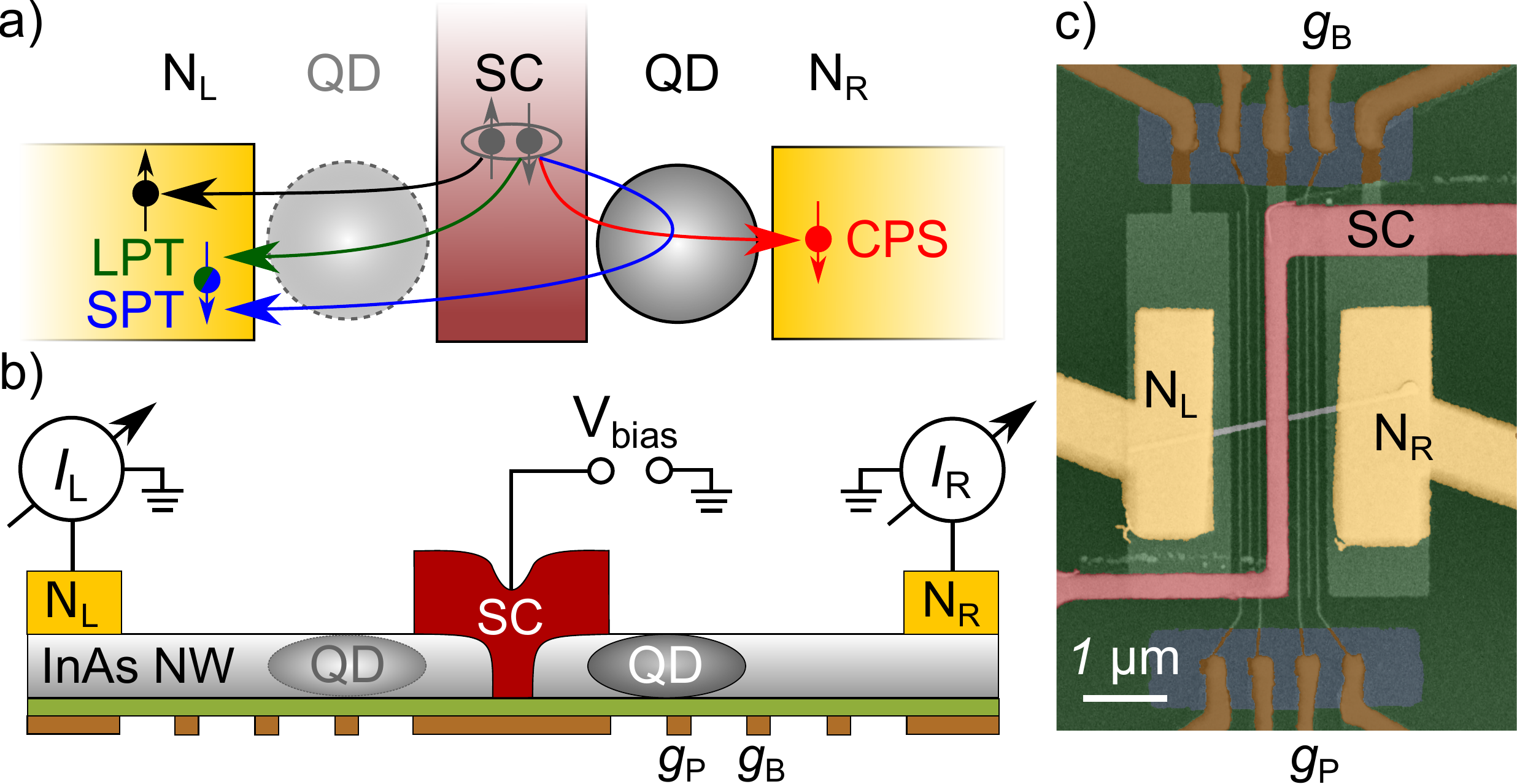}
	\caption{a) Schematics of a Cooper pair splitter. Cooper pairs are extracted from a central superconducting electrode (SC) to two normal leads (N$_{\text{L}}$ and N$_{\text{R}}$), through quantum dots (QDs). The currents are built up of two contributions, the Cooper pair splitting (CPS, in red) and the local pair tunneling (LPT, in blue). In the presence of a Shiba state -- if the SC and the QD are strongly coupled -- a third process, the Shiba-assisted local pair tunneling (SPT, in blue) emerges. This produces a CPS-like non-local signal but does not create spatially separated electron pairs. The left QD is drawn with a dashed line, because it is not tuned throughout this work, but fixed as it only weakly transmits.
	b-c) Cross-section and false color SEM image of the device. The InAs nanowire (NW) is placed on an array of bottom gate electrodes and contacted by the two normal and one superconducting electrode. The QD is formed in the NW, and tuned by the voltage on the plunger gate, $g_{\text{P}}$. The tunnel coupling to the N$_{\text{R}}$ lead is tuned by the barrier gate, $g_{\text{B}}$.}
	\label{fig1}
	\end{center}
	\end{figure}

Cooper pair splitter devices were extensively studied both experimentally \cite{HofstetterNature2009,HermannPRL2010,WeiNatPhys2010,HofstetterPRL2011,DasNatComm2012,SchindelePRL2012,TanPRL2015} and theoretically \cite{LesovikEPJB2001,RecherPRB2001,EldridgePRB2010,HiltscherPRB2011,LeijnsePRL2013,VeldhorstPRL2010, BursetPRB2011,SatoPRB2012, SollerBeilstein2012, CottetPRB2012,GiovannettiPRB2012} in recent years. They are proposed to serve as the source of spatially separated entangled electron pairs. The standard Cooper pair splitter (see Fig.~\ref{fig1}a) consists of a central superconducting electrode, which serves as the source of Cooper pairs, two normal leads, to which the split electron pair is transferred, and two quantum dots in between. The gate voltages of the quantum dots serve as the experimental knobs to tune the currents in the two different arms, utilizing the discrete energy spectrum of the dots. 

Theoretically, several ways were proposed to test the performance of a Cooper pair splitter device, starting from current or noise correlations \cite{LossPRL2000,ChevallierPRB2011,RechPRB2012}, through spin selective measurements \cite{LorenzoPRL2005,MortenEPL2008,SollerEPL2013,BrangePRL2017} to proving the entanglement by demonstrating the violation of the Bell-inequality \cite{ChtchelkatchevPRB2002,SamuelssonPRL2003,BednorzPRB2011,GiovannettiPRB2012,BrauneckerPRL2013,KlobusPRB2014,NiggPRB2015,BuszPRB2017} or using more complex spin-qubit or circuit quantum electrodynamics architectures \cite{CottetPRL2012,CottetPRB2012,ScherublPRB2014} or even time domain measurements \cite{BrangeArXiv2021}.

The experiments primarily focused on spatial current correlations \cite{HofstetterNature2009,HermannPRL2010,WeiNatPhys2010,HofstetterPRL2011,DasNatComm2012,SchindelePRL2012,TanPRL2015}. In general, the currents flowing in the two arms are attributed to two different processes. The first process is the \textit{local pair tunneling} (LPT), when both electrons of a Cooper pair are transferred to the same normal lead (see the black and green arrow on Fig.~\ref{fig1}a). This process involves only one arm of the device, correspondingly it is independent of control parameters of the other arm. The second process is the \text{Cooper pair splitting} (CPS), when the two electrons end up in different normal electrodes (see the black and red arrows on Fig.~\ref{fig1}a). As both arms of the device are involved, the CPS process depends on the control parameters of both arms. Hence it results in a \textit{non-local signal}, i.e. the current in one arm of the device is tuned by the control parameters (on-site energy/plunger gate voltage and tunnel amplitude / barrier gate voltage) of the other arm via tuning the CPS contribution.

In Ref.~\onlinecite{ScherublNatComm2020}, we have shown that a non-local signal can be also present when the conductance of one of the 
arms of the Cooper pair splitter is quenched. It is achieved by isolating the quantum dot from its normal lead by a strong tunnel barrier. Since in this situation the CPS is forbidden, there must be another transport process that produces a non-local signal. This process emerges if the tunnel coupling between the superconductor and the isolated quantum dot is sufficiently strong, which gives rise to the formation of the Yu-Shiba-Rusinov (Shiba) state \cite{YuActaPhysSin1965,ShibaProgTheorPhys1968,RusinovJETP1969,BalatskyRMP2006}.

The subgap Shiba state is the result of the hybridization of the dot states and quasiparticle states of the superconductor and was extensively studied in a variety of hybrid nano devices \cite{BuitelaarPRL2002,EichlerPRL2007,JespersenPRL2007,DeaconPRL2010,DeaconPRB2010,DirksNatPhys2011,KimPRL2013,PilletNatPhys2010,ChangPRL2013,PilletPRB2013,KumarPRB2014,SchindelePRB2014,LeeNatNano2014,JellinggaardPRB2016,LeePRB2017,GramichPRB2017,LiPRB2017,BretheauNatPhys2017,SuNatComm2017,SuPRL2018,SuPRB2020,EstradaSaldanaCommsPhys2020,EstradaSaldanaPRB2020,ValentiniArXiv2020,JungerPRL2020,GarciaCorralPRR2020}. They inherited their name from the subgap states in an analogous system, namely magnetic adatoms on a superconducting surface, which are widely studied by scanning tunneling microscope experiments \cite{YazdaniScience1997,JiPRL2008,HatterNatComm2015,MenardNatPhys2015,RubyPRL2016,ChoiNatComm2017,CornilsPRL2017,MenardNatComm2017,FarinacciPRL2018,HeinrichProgSurfSci2018,KezilebiekeNanoLett2018,SchneiderArXiv2019,KamlapureArXiv2019,PetersNatPhys2020,FarinacciPRL2020,BeckNatComm2021,WangPRL2021}. We will call the transport mechanism responsible for the non-local signal in Ref.~\onlinecite{ScherublNatComm2020} \textit{Shiba-assisted local pair tunneling} (SPT), since it is enabled by the Shiba state, and it involves subsequent transitions of two electrons (elements of a Cooper pair) into the same normal lead.

The experimental evidence of the SPT process, provided in Ref.~\onlinecite{ScherublNatComm2020}, implies that a non-local signal in a Cooper pair splitter with strong superconductor-dot coupling is induced {\it together} by CPS and SPT. In turn, this implies that in general it is unjustified to interpret the entire non-local signal as CPS, and extra efforts are required to distinguish the contributions of CPS and SPT to the non-local signal. Here, we exemplify and illustrate this challenge through experimental data, establish a theoretical framework allowing us to separate the two contributions, and show that the framework can indeed be successfully applied to our experiment, i.e. it separates the CPS and SPT contributions. 

In the experiment (see Fig.~\ref{fig1}a for the experimental setup, to be detailed below), we carried out transport measurements while the tunnel coupling between the quantum dot and the N$_{\text{R}}$ lead was varied. Our transport model is based on the so-called zero bandwidth approximation (ZBA) for the superconductor \cite{AffleckPRB2000,JellinggaardPRB2016,ProbstPRB2016,GroveRasmussenNatComm2018}. Through our analysis, we find that close to the quenched regions (i.e. when the tunnel barrier toward N$_\text{R}$ is strong), a significant contribution of the non-local signal comes from the SPT process.
Hence, in this parameter range, it is necessary to use our framework for a sound analysis. Since following earlier approaches, e.g. Refs.~\onlinecite{HofstetterNature2009,HermannPRL2010,DasNatComm2012,SchindelePRL2012,TanPRL2015}, attributing all of the non-local signal to CPS would overestimate the real CPS contribution in our case. We also show that in the case when the quantum dot of the Cooper pair splitter is well coupled to the N$_{\text{R}}$ lead, the non-local signal dominantly comes from the splitting of Cooper pairs, and correspondingly the earlier approaches gives a reasonable result for the CPS contribution in that parameter range.

Our paper is structured as follows. In Section~\ref{sec:device} the device geometry and measurement techniques are introduced. In Sec.~\ref{sec:exp} the experimental results are presented. We demonstrate the presence of a non-local signal for a wide range of parameters, we show a correlation between the local and the non-local signal and we relate them to the Shiba state. In Sec.~\ref{sec:model} the theoretical model and the transport calculation are introduced, and are used to fit the experimental data. In Sec.~\ref{sec:disc} we discuss our findings, namely, the origin of the non-local signal, the correlation between the local and the non-local signal, we calculate the contributions of the CPS and SPT processes, and discuss the validity of earlier splitting efficiency definitions. Finally in Sec.~\ref{sec:concl} we conclude our findings.

\section{Device geometry}

\label{sec:device}

The Cooper pair splitter device is illustrated in Fig.~1. The schematics of the cross-section and a false-color SEM image of the device is shown on panel b) and c), respectively. The basis of the circuit is an InAs nanowire (NW), with a diameter of 70~nm, grown by gold catalyst assisted molecular beam epitaxy \cite{AagesenNatNanotech2007,MadsenJCrystGrow2013}, using a two-step growth method to suppress the stacking faults \cite{ShtrikmanNanoLett2009}. First, an array of bottom gate electrodes (made of 4/18~nm Ti/Pt in brown) was defined by e-beam lithography and evaporation. The two outermost electrodes of the 9 are 1.3~$\mu$m wide and placed below the normal contacts. The middle electrode is 250~nm wide and placed below the superconductor. The remaining 3+3 electrodes are 30~nm wide with 100~nm period and serve to tune the electron density in the nanowire. The gate electrodes are covered by a 25~nm thick SiN$_\text{x}$ layer (green), grown by plasma-enhanced chemical vapor deposition to electrically isolate the gates from the rest of the device. The SiN$_\text{x}$ was removed by reactive ion etching with $\text{CHF}_3/\text{O}_2$ at the end of the gate electrodes to contact them \cite{WongJVacSciTechB1992}. The nanowire is contacted by two 4.5/100~nm thick Ti/Au normal electrodes (N$_\text{L}$ and N$_\text{R}$ in yellow) and a 4.5/110/20~nm thick Pd/Pb/In superconducting contact (SC in red) \cite{GramichAPL2016}. The nanowire was cut by focused ion beam prior to the deposition of the superconducting contact (see Fig.~\ref{fig1}b) to suppress the direct tunnel coupling between the two arms of the device \cite{FulopPRB2014,FulopPRL2015}.

The differential conductance of the two arms were measured simultaneously with lock-in technique, using 10~$\mu$V AC signal on the superconducting contact at 237~Hz, on the two normal contacts by home-built I/V converters. The measurements were carried out in a Leiden Cryogenics CF-400 top loading cryo-free dilution refrigerator at a bath temperature of 35~mK. Prior to cooldown, the sample was pumped overnight to remove the absorbed water from the surface of the nanowire.
	
\section{Experimental results}

\label{sec:exp}

\begin{figure*}[!htbp]
	\begin{center}
	\includegraphics[width=0.8\textwidth]{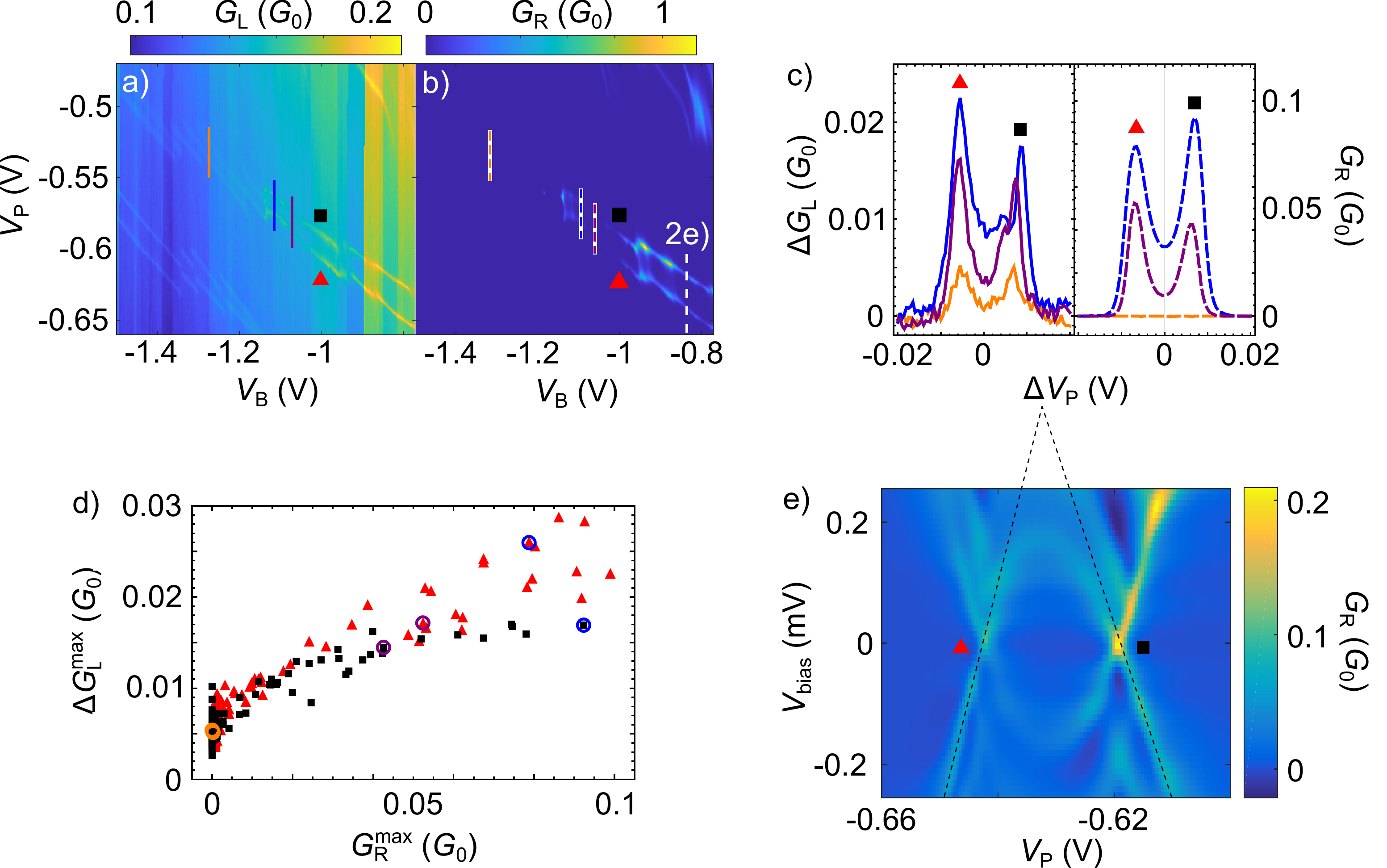}
	\caption{a,b) Simultaneously measured zero-bias differential conductances of the two arms of the device, as the functions of the $V_{\text{B}}$ (barrier) and $V_{\text{P}}$ (plunger) gate voltages. The quantum dot-like resonances of the right arm (marked with black square and red triangle in panel b) also appear in the conductance of the left arm (panel a). This non-local signal persists even when the conductance of the right arm is quenched. c) Color-coded line cuts from panels a) and b) illustrating the correlation between the non-local signal in the left arm and the local conductance of the right arm. A gate-dependent background conductance is subtracted from each $G_{\text{L}}$ curve. The conductance maxima values are extracted from each line cut along the resonances, as illustrated on panel c) by the marks. These value-pairs are plotted as a scatter plot in panel d) to further demonstrate the correlation of the two conductances. The three particular example pairs of panel c) are highlighted by color-coded circles. The non-local signal in the left arm increases with an increasing conductance of the right arm.
	e) Finite-bias characterization of the quantum dot with open barrier, measured along the white dashed line on panel b), at $V_{\text{B}}=-0.83$~V. The eye-shaped crossing is the signature of a Shiba state, which highlights the strong coupling between the superconductor and the quantum dot. The black dashed lines are used to estimate the charging energy to $U\approx 0.4$~meV.}
	\label{fig2}
	\end{center}
	\end{figure*}

The experimental results are summarized in Fig.~2. Panels a) and b) show the simultaneously measured zero-bias differential conductance (in units of $G_0 = 2e^2/h$) of the two arms of the Cooper pair splitter device, $G_{\text{L}}$ and $G_{\text{R}}$, as the function of the gate voltages, $V_{\text{B}}$ and $V_{\text{P}}$, which are applied on gate $g_{\text{B}}$ and $g_{\text{P}}$, respectively. Note that both gate electrodes are placed under the \textit{same} arm of the device (see Fig.~\ref{fig1}b), and the left arm is not tuned, but fixed as it weakly transmits. Hence, contrary to usual Cooper pair splitters, instead of a quantum dot and a normal lead, the left arm can be considered as a tunnel coupled normal electrode. The barrier gate voltage, $V_{\text{B}}$ dominantly tunes the tunnel coupling to the N$_{\text{R}}$ electrode, while $V_{\text{P}}$, the plunger gate voltage tunes the level position of the quantum dot. The conductance $G_\text{R}$ through the right arm (see Fig.~\ref{fig2}b) is completely quenched for $V_{\text{B}}<-1.2$~V, while for more positive barrier gate voltages it starts to conduct along two parallel resonance lines (marked by a square and a triangle). This conduction pattern indicates a quantum dot-like behavior.

Remarkably, the two $G_\text{R}$ resonances of the quantum dot also appear in the conductance $G_\text{L}$ of the left arm (see Fig.~\ref{fig2}a) as a conductance enhancement on top of a smooth background. We will denote this conductance enhancement by $\Delta G_\text{L}$, and call it the \textit{non-local signal}. When the conductance of the right arm is finite ($V_{\text{B}}>-1.1$~V), the presence of this non-local signal is in agreement with previous measurements on Cooper pair splitters \cite{HofstetterNature2009,HermannPRL2010,DasNatComm2012,SchindelePRL2012,TanPRL2015}. I.e., when the quantum dot in the right arm is tuned to resonance, an additional transport channel opens, Cooper pairs are not only transferred to the left normal electrode by LPT processes, but they can be split between the two arms, and correspondingly increasing the conductance of both arms. This $V_\text{B} > -1.1$~V regime of the
data is qualitatively consistent with the picture of CPS.

However, the $V_\text{B} < -1.2$~V range of the data in Fig.~\ref{fig2} a) and b) excludes the scenario that CPS is the only transport process inducing the non-local signal. There, the conductance $G_\text{R}$ of the right arm is suppressed, but the non-local signal, i.e., the peaks in $G_{\text{L}}$, persists throughout the whole map. When $G_{\text{R}}$ is quenched, the presence of the non-local signal cannot be attributed to the CPS anymore, since no electron can flow to the N$_{\text{R}}$ electrode. Both electrons of the Cooper pairs have to leave to the left normal lead.

Furthermore, our data reveals that the amplitude of the non-local signal $\Delta G_\text{L}$ is controlled via the local conductance $G_\text{R}$. In other words, by tuning the rightmost barrier more transparent, the current moving in the left arm is also increased. This is illustrated in panel~c) of Fig.~\ref{fig2}, by line cuts from panels~a) and b) along the color coded lines. The smooth background conductance is subtracted from each line cut of $G_{\text{L}}$ and only the non-local signal, $\Delta G_{\text{L}}$ is plotted. When $G_{\text{R}}$ is quenched, the non-local signal goes up to about $\Delta G_\text{L} \approx 0.005~G_0$ (see the orange curves on panel~c), measured at $V_{\text{B}}=-1.281$~V). As the local conductance increases, the amplitude of the non-local signal also increases (see the purple and blue curves, measured at $V_{\text{B}}=-1.098$~V and $-1.119$~V, respectively). Note that $G_{\text{R}}$, and correspondingly the non-local signal also, varies in a non-monotonic fashion as the barrier gate voltage is tuned; we attribute this to disorder in the sample.

The above-mentioned systematic relation of the local conductance $G_\text{R}$ and the non-local signal $\Delta G_\text{L}$ is reinforced by the data shown in Fig.~\ref{fig2}d. Let us exemplify how we derive this data from the previous panels, using the line cuts in Fig.~\ref{fig2}c, focusing on the right maxima of the blue curves (marked by black squares). The maximum value of the non-local signal (left panel) is about $\Delta G^{\text{max}}_{\text{L}} \approx 0.017~G_0$, while the maximum value of the local conductance (right panel) is $G^{\text{max}}_{\text{R}} = 0.08~G_0$. These values give the vertical and horizontal coordinates of the corresponding data point in panel~d), respectively (see the black square highlighted by the blue circle). The other points of panel~d) were obtained similarly by reading the maxima of all vertical line cuts on panels~a) and b). As Fig.~\ref{fig2}d shows, even when the conductance of the right arm is quenched, i.e. $G^{\text{max}}_{\text{R}}=0$, the amplitude of the non-local signal is finite, residing between $0.002~G_0$ and $0.01~G_0$. As $G_{\text{R}}$ is increased by lowering the barrier to the N$_{\text{R}}$ lead, the non-local signal also increases in a monotonous, saturating fashion. Only data points with $G^{\text{max}}_{\text{R}}<0.1~G_0$ are shown in the figure. We choose this requirement to ensure the weak coupling to the right lead, which will be assumed in the modeling.

Using finite-bias measurements, we establish that the tunnel coupling between the superconducting lead and the quantum dot is strong, and hence a Shiba state is formed in the device. The finite-bias differential conductance measured along the white dashed line on Fig.~\ref{fig2}b) at $V_{\text{B}}=-0.83$~V is shown in Fig.~\ref{fig2}e). The eye-shaped crossing conductance lines are the usual signature of a Shiba state, formed in a strongly coupled superconductor--quantum dot hybrid (see e.g. Ref.~\onlinecite{LeeNatNano2014}). Correspondingly, the resonances on panels~a) and b) of Fig.~\ref{fig2} are the signatures of the zero-energy Shiba states rather than charge degeneracies of a quantum dot. These observations suggest that in our model of the device (see next section) we have to incorporate a strong, coherent hybridization between the superconductor and the quantum dot. Based on the finite-bias measurement shown on panel e) the charging energy of the quantum dot can also be estimated, following the dashed lines, yielding $U\approx0.4$~meV.

Our conclusion from this analysis is that when the conductance $G_\text{R}$ of the right arm is quenched, the conductance enhancement $\Delta G_\text{L}$ of the left arm originates from the (zero-energy) Shiba state. i.e. the SPT mechanism, rather than the splitting of Cooper pairs. Hence, the conductance enhancement can be viewed as the non-local probing of Shiba state \cite{ScherublNatComm2020}.

In the following, we will use a simple model to explain the experimental findings, namely i) the presence of the non-local signal when the right arm is quenched and ii) the correlation between the conductances when $G_{\text{R}}$ is finite. We will model the Shiba state using the zero bandwidth approximation for the superconductor. This model will enable us to identify and separate contributions of the SPT and CPS processes (see Fig.~\ref{fig1}a) to the non-local signal $\Delta G_\text{L}$.
	
\section{Model}

\label{sec:model}

\subsection{Zero-bandwidth approximation of the Shiba state}

The Shiba state is formed by the hybridization of the quasiparticles of the superconductor and the quantum dot degrees of freedom induced by the strong tunnel coupling between the dot and the superconductor. A simple way to describe the Shiba state is by treating the superconductor using the zero bandwidth approximation (ZBA) \cite{AffleckPRB2000,JellinggaardPRB2016,ProbstPRB2016,GroveRasmussenNatComm2018}. In this approximation, the quasiparticles of the superconductor are restricted only to a discrete energy level at energy $\Delta$, which can host 0, 1 or 2 quasiparticles. The corresponding Hamiltonian is
\bnen H^{\text{ZBA}}_{\text{SC}} = \Delta c^{\dagger}_{\uparrow} c^{\dagger}_{\downarrow} + \Delta c_{\downarrow} c_{\uparrow} = \Delta \sum_{\sigma} \gamma^{\dagger}_{\sigma} \gamma_{\sigma}, \eden
where $\Delta$ is the superconducting gap, $c^{(\dagger)}_{\sigma}$ is the electron annihilation (creation) operator in the superconductor and $\gamma^{(\dagger)}_{\sigma}$ is the annihilation (creation) operator of the quasiparticles, defined by the usual Bogoliubov-transformation, $c_{\sigma} = \frac{1}{\sqrt{2}} \left( \gamma_{\sigma} - \sigma \gamma^{\dagger}_{\bar\sigma} \right)$.

\begin{figure*}[!htbp]
	\begin{center}
	\includegraphics[width=\textwidth]{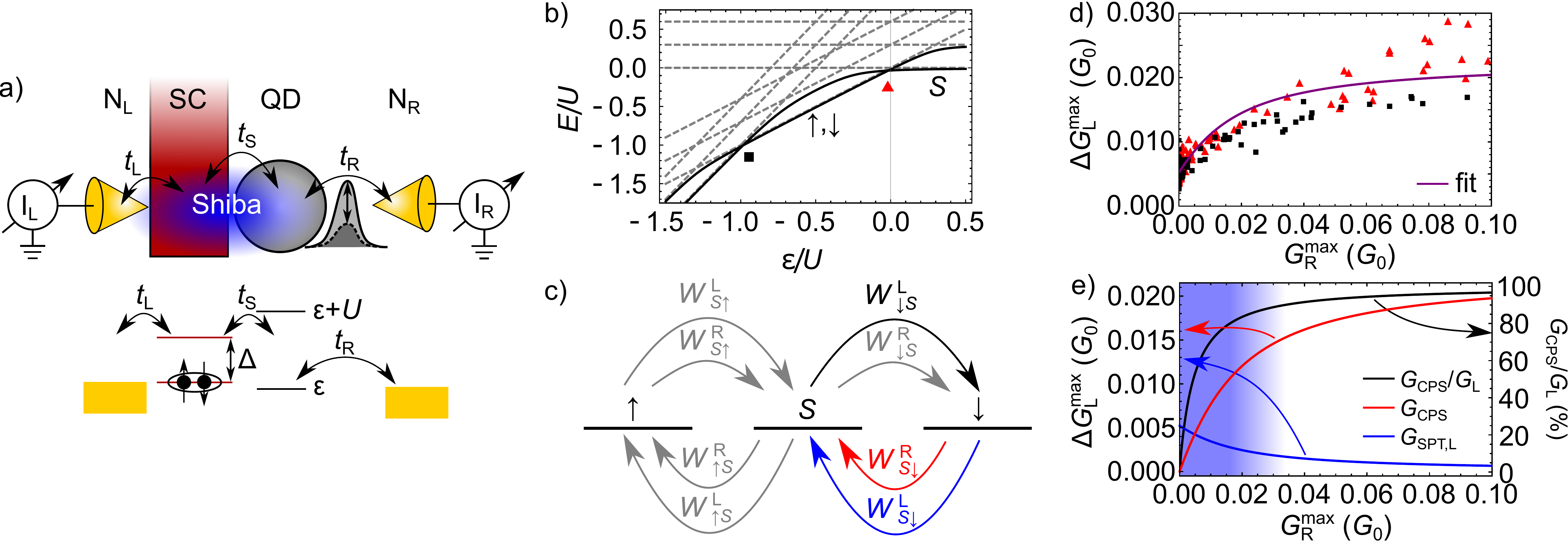}
	\caption{a) Schematics (top) and energy diagram (bottom) of our model. The tunnel coupling $t_{\text{S}}$ between the superconductor (SC) and quantum dot (QD) creates a Shiba state (in blue). The normal leads, N$_{\text{L}}$ and N$_{\text{R}}$ probe the Shiba state with coupling strength $t_{\text{L}}$ and $t_{\text{R}}$, respectively. For simplicity, the SC is treated within the zero bandwidth approximation, in which the quasiparticles are restricted to a discrete energy level at $\Delta$ energy. 
	b) Illustrative energy spectrum of the QD--SC hybrid as the function of the dot's level position $\varepsilon$. Gray dashed lines are the energies of the uncoupled system, while the black continuous lines are the possible ground states of the coupled system, the \textit{singlet} (\textit{S}) and the \textit{doublets}, $(\uparrow,\downarrow)$. The ground state transitions are marked with red triangle and the black square in accordance with Fig.~\ref{fig2}.
	c) Illustration of the dynamics induced by the coupling to the normal leads. The tunneling induces transitions between the different ground states of the dot--superconductor system, the singlet and the doublets. The highlighted, colored arrows illustrate the tunnel processes shown on Fig.~\ref{fig1}a).
	d) Fitting the experimental data of Fig.~\ref{fig2}d) with the ZBA model. Purple line is the fit with $t_{\text{S}}=0.145$~meV, $\Gamma_{\text{L}}=0.24~\mu$eV and $\Gamma_{\text{R}}=0-0.89~\mu$eV.
	e) Contribution of the CPS (red) and the SPT (blue) processes to the non-local signal of the left arm and the relative ratio of the CPS processes (black). In the shaded blue region the non-local signal is not a good measure of the CPS, since the SPT contribution is not negligible.}
	\label{fig3}
	\end{center}
	\end{figure*}

The quantum dot is described by the single-impurity Anderson Hamiltonian,
\bnen \label{eq:QD} H_{\text{QD}} = \sum_{\sigma} \varepsilon n_{\sigma} + U n_{\uparrow} n_{\downarrow}, \eden
where $\varepsilon$ is the level position of the dot, $U$ is the charging energy and $n_{\sigma}=d^{\dagger}_{\sigma} d_{\sigma}$ is the number of the electrons with spin $\sigma$ on the dot, with $d^{(\dagger)}_{\sigma}$ being the annihilation (creation) operator of electrons with spin $\sigma$ on the dot. Note that compared to the usual Cooper pair splitter geometry, where quantum dots are placed on both side of the superconductor (as it is illustrated in Fig.~\ref{fig1}a), in this model we only consider the dot of the right arm.
Since in the experiment the left arm of the device was not tuned, but fixed in weakly transmitting region, the left normal electrode and the left dot is modeled as a tunnel coupled normal lead for simplicity (see Fig.~\ref{fig3}a) and Sec.~\ref{sec:transport}).

The tunnel coupling between the superconductor and the dot is described by
\bnen H_{\text{TS}} = t_{\text{S}} \sum_{\sigma} \left( d^{\dagger}_{\sigma} c_{\sigma} + c^{\dagger}_{\sigma} d_{\sigma} \right). \eden
Using the Bogoliubov transformation above, this translates to
\bnen \label{eq:ZBA2} H_{\text{TS}} = \frac{1}{\sqrt{2}}\sum_{\sigma} t_{\text{S}} \left[ d^{\dagger}_{\sigma} \left( \gamma_{\sigma} - \sigma \gamma^{\dagger}_{\bar\sigma} \right) + \left( \gamma^{\dagger}_{\sigma} - \sigma \gamma_{\bar\sigma} \right) d_{\sigma} \right]. \eden

The composite system of the superconductor and the quantum dot is hence modelled by $H_\text{SC}^\text{ZBA} + H_\text{QD} + H_\text{TS}$. A natural basis of the corresponding 16-dimensional Fock space is the product basis of the electron Fock space of the QD and the quasiparticle Fock space of the superconductor, that is, the basis $\ket{i,j}=\ket{i}_{\text{QD}}\otimes\ket{j}_{\text{SC}}$. Here,  $\left\{\ket{0}_{\text{QD}},\ket{\uparrow}_{\text{QD}},\ket{\downarrow}_{\text{QD}},\ket{\uparrow\downarrow}_{\text{QD}}\right\}$ is the canonical basis of the electron Fock space of the QD and $\left\{\ket{0}_{\text{SC}},\ket{\uparrow}_{\text{SC}},\ket{\downarrow}_{\text{SC}},\ket{\uparrow\downarrow}_{\text{SC}}\right\}$ is the canonical basis of the quasiparticle Fock space of the superconductor. The basis states are eigenstates of the uncoupled QD-superconductor system, i.e., of $H_\text{SC}^\text{ZBA} + H_\text{QD}$. The energy spectrum of the uncoupled system is illustrated on Fig.~\ref{fig3}b with gray dashed lines as the function of the dot's level position, $\varepsilon$. The three different slopes correspond to the different fillings of the dot, while from the parallel lines the bottom one corresponds to zero, the middle one to one and the top one to two quasiparticles in the superconductor.
The product basis can be partitioned to states with even fermion-number parity and odd fermion-number parity, the two subset of basis states spanning the even-parity and odd-parity sectors (subspaces), respectively.

The tunnel coupling $H_\text{TS}$ hybridizes the states within each parity sector in two ways. First, it converts a quasiparticle from the superconductor to an electron on the dot (or the other way around), and second, it splits a Cooper pair to an electron on the dot and to a quasiparticle in the superconductor (or the other way around). Note that although the latter process does preserve the fermion-number parity, it does not preserve the fermion number.  
 
The eigenstates of the coupled superconductor--quantum dot system can be grouped into six invariant subspaces, which can be labeled by the total spin of the electrons and the quasiparticles. The five-dimensional spinless, \textit{singlet} (\textit{S}) sector is spanned by the states $\left\{\ket{0,0},\ket{\uparrow\downarrow,0},\ket{0,\uparrow\downarrow},\ket{\uparrow\downarrow,\uparrow\downarrow}\right\}$, and the state $\frac{1}{\sqrt{2}}\left(\ket{\uparrow,\downarrow}-\ket{\downarrow,\uparrow}\right)$. In the absence of magnetic field, the 8 states with odd number of fermions can be grouped into two subspaces. The states $\left\{\ket{\uparrow,0},\ket{0,\uparrow},\ket{\uparrow\downarrow,\uparrow},\ket{\uparrow,\uparrow\downarrow}\right\}$ $\left(\left\{\ket{\downarrow,0},\ket{0,\downarrow},\ket{\uparrow\downarrow,\downarrow},\ket{\downarrow,\uparrow\downarrow}\right\}\right)$ form the two-fold degenerate \textit{doublet} (\textit{D}) subspaces. These states have $+1/2$ ($-1/2$) spin $z$ component. The remaining three triplet states are uncoupled from the other states and from each other.

To model our Shiba-related transport measurement results, only the lowest-energy singlet and doublet states are required. The former one (latter ones) will be denoted by $\ket{S}$ ($\ket{\sigma}$, with $\sigma \in \{\uparrow,\downarrow\}$). These states are illustrated on Fig.~\ref{fig3}b as black lines.

\subsection{Transport calculation}

\label{sec:transport}

In the experiment, the system is simultaneously probed by two tunnel-coupled normal electrodes, as illustrated in Fig.~\ref{fig3}a. The tunnel coupling to the normal leads are described by
\bean H_{\text{TL}} &=& t_{\text{L}} \sum_{\sigma} \left( c^{\dagger}_{L\sigma} c_{\sigma} + c^{\dagger}_{\sigma} c_{L\sigma} \right) \nonumber \\
&=& \frac{t_{\text{L}}}{\sqrt{2}} \sum_{\sigma }\left[ c^{\dagger}_{L\sigma} \left( \gamma_{\sigma} - \sigma \gamma^{\dagger}_{\bar\sigma} \right) + \left( \gamma^{\dagger}_{\sigma} - \sigma \gamma_{\bar\sigma} \right) c_{L\sigma} \right] \nonumber \\
 H_{\text{TR}} &=& t_{\text{R}} \sum_{\sigma} \left( c^{\dagger}_{R\sigma} d_{\sigma} + d^{\dagger}_{\sigma} c_{R\sigma} \right), \eean
where $c^{(\dagger)}_{L/R\sigma}$ are the annihilation (creation) operators of the electrons in the left/right normal leads. The left lead couples only to the superconductor, while the right one only to the quantum dot. 

As it was stated above, the Hamiltonian of the superconductor--quantum dot system, $H_\text{SC}^\text{ZBA} + H_\text{QD} + H_\text{TS}$ preserves the parity of the system, while the tunnel coupling to the normal leads induces transitions between eigenstates with different parity while transferring individual electrons to the normal leads. These transitions are illustrated in Fig.~\ref{fig3}c and described in the following.

For simplicity, we assume zero temperature, a gate-voltage configuration enabling resonant tunneling, i.e. that $E_S = E_\sigma$ for the energy eigenvalues of the singlet and double ground states, and that a small negative bias voltage is applied on the superconducting lead, to ensure that electrons flow only towards the normal leads. Note that the resonant-tunneling condition is satisfied for two different on-site energies, in the vicinities of $\varepsilon = 0$ and $\varepsilon = -U$, respectively, and our approach described here is valid for both configurations.

Since the tunnel couplings to the normal leads were kept weak in the experiment, we treated the electron transport to the normal leads perturbatively, using Fermi's golden rule. The non-zero transition rates are
\bean \label{eq:rates} W^{\text{L}}_{S\sigma} &=& \frac{\Gamma_{\text{L}}}{\hbar} \left| \left\langle S \left| \frac{\gamma_{\sigma} - \sigma \gamma^{\dagger}_{\bar\sigma}}{\sqrt{2}} \right| \sigma \right\rangle \right|^2, \nonumber \\
W^{\text{L}}_{\sigma S} &=& \frac{\Gamma_{\text{L}}}{\hbar} \left| \left\langle \sigma \left| \frac{\gamma_{\bar\sigma} - \sigma \gamma^{\dagger}_{\sigma}}{\sqrt{2}} \right| S \right\rangle \right|^2, \nonumber \\
W^{\text{R}}_{S\sigma} &=& \frac{\Gamma_{\text{R}}}{\hbar} \left| \bra{S} d_{\sigma} \ket{\sigma} \right|^2, \nonumber \\
W^{\text{R}}_{\sigma S} &=& \frac{\Gamma_{\text{R}}}{\hbar} \left| \bra{\sigma} d_{\bar\sigma} \ket{S} \right|^2, \eean
where $\Gamma_{\text{L/R}} = \pi \rho_{\text{L/R}} t^2_{\text{L/R}}$ is the coupling to the left/right normal lead with $\rho_{\text{L/R}}$ being the density of state of lead N$_{\text{L/R}} $. The \textit{f/i} indices of $W^{\text{L/R}}_{fi}$ denote the \textit{final/initial} states of the given transition. The transition matrix elements are calculated at the $\varepsilon$ values, where the singlet and doublet ground states are degenerate as in the measurement, outlined in Sec.~\ref{sec:exp}.

The tunnel coupling to the normal leads induces parity changing transitions between the different states of the superconductor--quantum dot hybrid as it is illustrated in Fig.~\ref{fig3}c. For example, the black arrow denotes the tunneling event, which brings the system from the $\ket{S}$ state to the $\ket{\downarrow}$ state, while an electron is transferred to the N$_{\text{L}}$ lead. The time evolution of the occupation probabilities of the $\ket{S/\sigma}$ states, $P_{S/\sigma}$ is described by a classical master equation,
\bean \label{eq:MasterEq} \frac{dP_S}{dt} &=& \sum_{\sigma} \left[ \left(W^{\text{L}}_{S\sigma} + W^{\text{R}}_{S\sigma} \right) P_{\sigma} - \left(W^{\text{L}}_{\sigma S} + W^{\text{R}}_{\sigma S} \right) P_S \right] \nonumber \\
 \frac{dP_{\sigma}}{dt} &=& \left(W^{\text{L}}_{\sigma S} + W^{\text{R}}_{\sigma S} \right) P_S - \left(W^{\text{L}}_{S\sigma} + W^{\text{R}}_{S\sigma} \right) P_{\sigma}, \eean
together with the normalization condition $P_S+ \sum_{\sigma} P_{\sigma} =1$.

After determining the stationary solution ($dP_i/dt=0$) of the master equation, the current in lead N$_{\text{L/R}}$ is given by
\bnen \label{eq:curr} I_{\text{L/R}} = e \sum_{\sigma} \left( W^{\text{L/R}}_{\sigma S} P_S + W^{\text{L/R}}_{S\sigma} P_{\sigma} \right). \eden

To compare the model with the experiment, we assume a linear relation between the differential conductances and the currents, and use the $10~\mu$V AC voltage as the conversion factor,  i.e. $G_{\text{L/R}}=I_{\text{L/R}}/10~\mu\text{V}$. The validity of this approximation will be discussed below.

\subsection{Results of the fitting}

We used the framework detailed above to fit our experimental data shown in Fig.~\ref{fig2}d. The charging energy and the superconducting gap is fixed to the experimental value of $U=0.4$~meV and $\Delta=0.25$~meV, respectively. As $\varepsilon$ is fixed by the degeneracy condition, only $t_{\text{S}}$ remains a free parameter in the model of the coupled superconductor and QD. In the following we will focus first on the degeneracy point near $\varepsilon = 0$, i.e. the $0-1$ transition of the dot. Our results can be directly translated to the other degeneracy point, using symmetry arguments (it will be discussed below). Besides $t_{\text{S}}$, which determines the transition matrix elements, the couplings to the normal leads, $\Gamma_{\text{L}}$ and $\Gamma_{\text{R}}$ are also fitting parameters of the model. Note that $\Gamma_{\text{R}}$ is tuned in the experiment, therefore the fitting does not give a single value, but a range starting from zero, therefore the fitting parameter is not $\Gamma_\text{R}$ itself, but its maximal value $\Gamma^\text{max}_\text{R}$.

The fitted curve is shown in Fig.~\ref{fig3}d with the purple line, using $t_{\text{S}}=0.145$~meV, $\Gamma_{\text{L}}=0.24~\mu$eV and $\Gamma_{\text{R}}=0-0.89~\mu$eV. The fit captures well the main features of the experiment. First, the model reproduces the finite current in the left lead even when the conductance of the right arm is quenched. Second, $G_{\text{L}}$ also increases as $G_{\text{R}}$ is enhanced by the increasing $\Gamma_{\text{R}}$. And third, the model gives a saturating tendency for the non-local signal as the barrier to the right lead is opened up.

We remark that the $\Gamma_{\text{L}}$ and $\Gamma_{\text{R}}$ values obtained above by fitting the zero-bias differential conductance data underestimate the physical coupling strength in the device. The model calculation results in a zero-width conductance peak, with the peak height, and hence the DC current at high bias determined by the $\Gamma$ parameters. However, in the experiment the conductance peak has a finite width, which is probed at zero bias, with an AC voltage smaller than the peak width. We estimate the saturated current at high bias in the following manner. Using the FWHM of the gate-dependent resonance curve (see e.g. Fig.~\ref{fig2}c) and the lever arm of the plunger gate we determine the broadening of the resonance, yielding about 50~$\mu$eV. In the experiment we applied an AC voltage of 10~$\mu$V to probe the conductance, with extrapolation the saturated DC current is about ten times higher then the value used in the fitting above. Therefore, by comparing the saturated values we obtain $\Gamma_{\text{L}}=2.4~\mu$eV and $\Gamma_{\text{R}}=0-8.9~\mu$eV.

\section{Discussion}

\label{sec:disc}

Previously we demonstrated that the ZBA model of the Shiba state well reproduces our experimental findings: i) the presence of the non-local signal, even when $G_{\text{R}}$ is quenched and ii) the correlation between the height of the local and the non-local conductance peaks. The model allows for the detailed understanding of the experimental features and furthermore, it can distinguish between the SPT and CPS contribution of the currents.

First, we have to discuss why the Shiba language (i.e. treating the superconductor--quantum dot hybrid as a coherently coupled unit) is necessary to describe our measurements. Often in Cooper pair splitters the electron transition between the superconductor and quantum dots is also described by tunnel rates, derived in a perturbative model, which assumes weak tunnel couplings \cite{HofstetterNature2009,HermannPRL2010,SchindelePRL2012}. In our measurements the presence of the eye-shaped subgap resonances in Fig.~\ref{fig2}e implies that the tunnel coupling between the superconductor and the quantum dot is comparable to the value of the gap (which we also verified by the fit). Therefore it cannot be treated perturbatively, but the coherent hybridization of the superconductor and the dot has to be taken into account. Considering the usual, perturbative description of the Cooper pair splitters, the presence of the non-local signal in the region, where $G_{\text{R}}$ is zero is surprising, since the process, which can generate a non-local signal -- the CPS -- is forbidden. However, in the Shiba description this non-local signal is naturally present, as we will see below.

Second, we observed that the amplitude of the non-local signal increases as $G_{\text{R}}$ is increased by enhancing the tunnel coupling to the N$_{\text{R}}$ lead. This correlation is a result of a bottleneck effect. When the right lead is isolated from the rest of the device, all electrons have to leave to the N$_{\text{L}}$ lead. This means that both the \textit{singlet}$\rightarrow$\textit{doublet} and the \textit{doublet}$\rightarrow$\textit{singlet} transitions must occur involving a tunneling to the left lead. If the transition rates are much different, the slower one creates a bottleneck. Indeed, using the model parameters obtained by the fit, one finds that $\hbar W^{\text{L}}_{S\sigma}\approx 0.009~\mu$eV (the blue and the top left gray arrow on Fig.~\ref{fig3}c) and $\hbar W^{\text{L}}_{\sigma S} \approx 0.048~\mu$eV (the black and the bottom left gray arrow on Fig.~\ref{fig3}c). The former rate is more than 5 times smaller than the latter one, i.e. the left current is limited by the \textit{doublet}$\rightarrow$\textit{singlet} transition. However, opening up the barrier to the right lead opens up another transport channel for the slower process. As the $W^{\text{R}}_{S\sigma}$ rate becomes larger than $W^{\text{L}}_{S\sigma}$, the \textit{doublet}$\rightarrow$\textit{singlet} process happens more favorably by transferring an electron to the right lead, and the bottleneck starts to be resolved. Although the presence of the new transport channel first means that not all the electrons are transferred to the left lead (which would mean the decrease of the non-local signal), but the whole transport sequence speeds up so much that altogether the left current increases.

Above we mentioned that during the fitting procedure only the $\varepsilon \lesssim 0$ degeneracy point was considered, because the results obtained there can be directly translated to the other degeneracy point at $\varepsilon \gtrsim -U$, corresponding to the $1-2$ transition of the dot. The data points marked with red triangles on Figs.~\ref{fig2}d and \ref{fig3}c were indeed measured at the former degeneracy point, but the black squares correspond to the latter one. Generally, the eigenstates are different combinations of the basis states at the two degeneracy points, therefore the connection between them is not straightforward. However, the fact that the two data sets run together (see Fig.~\ref{fig2}d) already foreshadows the relation between the two degeneracy points. For example the weight of the $\ket{0,x}$ state at the $\varepsilon \lesssim 0$ degeneracy point is the same as the weight of the $\ket{\uparrow\downarrow,x}$ at the $\varepsilon \gtrsim -U$ point, and the weight of the $\ket{\sigma,x}$ is the same at the two degeneracy points. It follows that the role of transition rates are reversed, e.g. at the $\varepsilon \gtrsim -U$ degeneracy point $\hbar W^{\text{L}}_{\sigma S}\approx 0.009~\mu$eV and $\hbar W^{\text{L}}_{S \sigma} \approx 0.048~\mu$eV (c.f. the values above). Therefore not the \textit{doublet}$\rightarrow$\textit{singlet}, but the \textit{singlet}$\rightarrow$\textit{doublet} transition gives the bottleneck.

The presence of non-local signals was reported previously in Cooper pair splitters hosting superconducting subgap states \cite{SchindelePRB2014,GramichPRB2017}. The authors used the infinite superconducting gap limit description -- frequently referred as Andreev bound states -- to model their experimental results. Using this approximation all transition matrix elements contain the $d$-operators' matrix elements, therefore the ratio of the left rates and the right rates are the same, i.e. $W^{\text{L}}_{\sigma S}/W^{\text{L}}_{S\sigma}=W^{\text{R}}_{\sigma S}/W^{\text{R}}_{S\sigma}$. With a simple derivation it can be shown that increasing $\Gamma_{\text{R}}$ cannot resolve the bottleneck effect, since it drops out of the formula of $I_{\text{L}}$. This implies that the infinite gap model cannot produce a correlation between the currents. Therefore, considering both the dot- and quasiparticle-components in the Shiba state is crucial to reproduce the dependence of the non-local signal on $\Gamma_{\text{R}}$. Note that in the previously reported experiments the tunnel couplings were not tuned systematically, therefore the infinite gap limit was sufficient for the explanation of the experimental findings \cite{SchindelePRB2014,GramichPRB2017}. 

In the literature, the non-local signals were purely attributed to the CPS process \cite{HofstetterNature2009,HermannPRL2010,DasNatComm2012,SchindelePRL2012,TanPRL2015}, however above we demonstrated that in the presence of the Shiba state the non-local signal is also present when the CPS is forbidden. It raises the question: what is the real contribution of the CPS process to the non-local signal? Or, alternatively, to what extent can we use the non-local signal as the measure of the CPS processes? The ZBA model allows us to answer these questions by calculating currents carried by the SPT and CPS processes separately. Here, first we introduce how these processes are defined based on single tunneling events, then we calculate their contribution.

The whole transport consists of a sequence of \textit{singlet}$\rightarrow$\textit{doublet} and \textit{doublet}$\rightarrow$\textit{singlet} transitions. Assuming that one starts from the singlet ground state the transport can be grouped into a random sequence of SPT and CPS processes. Let us illustrate this using panel c) of Fig.~\ref{fig3}. The highlighted arrows indicate a CPS and a SPT process. Starting from the singlet ground state, let us assume that first an electron (with up spin) is transferred to the left lead. This is indicated by the black arrow. At the end of the tunneling event the superconductor--quantum dot hybrid system stays in the $\ket{\downarrow}$ state. To return to the singlet state there are two possibilities. First, a second electron (with down spin) leaves to the left lead also (indicated by the blue arrow), and second, this electron is transferred to the right lead (red arrow). In the first process two electrons with opposite spins are transferred to the \textit{same} lead, therefore a SPT event occurred, while in the second process the two electrons end up in  \textit{different} leads, corresponding to a CPS process. As the system returned to the singlet state, the transport cycle starts again.

The contribution of the CPS and SPT processes can be calculated within the framework of our model. Substituting the stationary solution of the master equation \eqref{eq:MasterEq} to the expression of the currents, Eq.~\eqref{eq:curr} one finds that the currents can be partitioned as
\bnen \label{eq:part1} I_{\text{L/R}} = I_{\text{SPT,L/R}} + I_{\text{CPS}} \eden
where
\bnen \label{eq:part2} I_{\text{SPT,L/R}} = A \sum_{\sigma} 2 W^{{\text{L/R}}}_{S\sigma} W^{{\text{L/R}}}_{\sigma S} \eden
is the SPT contribution and
\bnen \label{eq:part3} I_{\text{CPS}} = A \sum_{\sigma} \left( W^{{\text{L}}}_{S\sigma} W^{{\text{R}}}_{\sigma S} + W^{{\text{R}}}_{S\sigma} W^{{\text{L}}}_{\sigma S}\right) \eden
is the CPS one. In $I_{\text{SPT,L/R}}$ the factor of 2 indicates that both electrons are transferred to the same lead, and the normalization factor is $A = 4e\left[\sum_{\sigma}\left(W^{{\text{L}}}_{S\sigma}+W^{{\text{R}}}_{S\sigma}+2W^{{\text{L}}}_{\sigma S}+2W^{{\text{R}}}_{\sigma S}\right)\right]^{-1}$.
Note that to obtain these formulas, the spin rotational symmetry was used, which implies the spin-independence of the transition rates, e.g. $W^{{\text{L}}}_{S\uparrow}=W^{{\text{L}}}_{S\downarrow}$. We should remind the reader that the presented model does not account for the LPT process, as it is independent of the Shiba state, consisting only of Cooper pairs tunneling from the superconductor directly to the normal lead. 

This separation of $I_{\text{L}}$ using the obtained fit parameters is plotted in Fig.~\ref{fig3}e. The CPS and SPT contributions are plotted in red and blue, respectively (The total current is shown on panel c) in purple). When $G_{\text{R}}=0$ the splitting of Cooper pairs is forbidden, since no electrons can be transferred to the right lead. Accordingly, the CPS contribution is zero. The total current in the left lead comes only from the SPT process. This limit corresponds to the non-local spectroscopy of the Shiba state ($V_\text{B}<-1.2$~V region on Fig.~\ref{fig2}a). As the conductance of the right arm is increased, the amplitude of the SPT processes decreases while the CPS one strongly increases. When the barrier to the right lead is transparent, about 95\% of the non-local signal comes from CPS processes. This is the usual Cooper pair splitter limit ($V_\text{B}>-1.1$~V region on Figs.~\ref{fig2}a\&b).
In between the two limits there is a region where a significant contribution of the non-local signal comes from SPT processes, this is indicated by the blue shaded region. In this region the usual splitting efficiencies, which assume that the non-local signal comes only from the splitting of Cooper pairs, overestimate the real contribution of CPS processes.

Our results can be generalized to the conventional Cooper pair splitter geometry, i.e. the N--QD--SC--QD--N setup in a straightforward way. The strongly coupled QD--SC--QD system hosting Andreev- or Shiba-molecular states was recently studied in Refs.~\onlinecite{ScherublBJNano2019} and \onlinecite{kurtossyArXiv2021}. Similarly to the presented model, in these systems the ground state can be either a unique singlet or a two-fold degenerate doublet and currents can flow to the normal leads when all of these three states are degenerate. Therefore, the transport processes illustrated on Fig.~\ref{fig3}c and the partitioning of the currents, presented in Eqs.~\eqref{eq:part1}-\eqref{eq:part3} also applies to the conventional splitter setup. The transport data measured in such a setup can be evaluated analogously to obtain the microscopic parameters and to partition the currents, only the transition matrix elements have to derived from the N--QD--SC--QD--N setup's Hamiltonian.

In conclusion, close to the quenched regime a significant contribution of the non-local signal comes from the SPT processes, but in the transparent barrier limit the non-local signal provides a good measure of the CPS processes.

\section{Conclusions}

\label{sec:concl}

We have investigated the non-local signal of a Cooper pair splitter as the tunnel coupling to one of the normal leads was varied. It was shown that the non-local signal persists even upon quenching the conductance in the opposite arm of the splitter, contradicting the expectations since in this limit the CPS process is forbidden. We resolved the contradiction by considering the Shiba state, originating from the hybridization of the quantum dot and the superconductor states. In the quenched limit, the non-local signal is attributed to the largely extended zero-energy Shiba state, which we studied in detail in Ref.~\onlinecite{ScherublNatComm2020}.

The unexpected non-local signal we observed implied that figures of merit for Cooper pair splitting efficiencies used in previous experimental works could not be applied for our device. Therefore, we constructed a transport model using the ZBA for the superconductor to reproduce our experimental findings and to separate the Shiba related non-local signal from the CPS. Despite of its simplicity, the model captures well the main features of the experiments. Utilizing the model and the obtained microscopic parameters we showed that closed to the quenched region, the non-local signal mostly consists of the SPT process, therefore, the previously applied efficiency indicators overestimate the real CPS contribution. However, when the tuned barrier is transparent the non-local signal is dominated by the CPS process, and accordingly, our results are consistent with those based on the standard CPS efficiency indicators.  

The data that underlies the plots within this paper and other findings of this study are available at \url{https://doi.org/10.5281/zenodo.5281979}.

\acknowledgments

We acknowledge Morten~H.~Madsen for MBE growth, Titusz~Feh\'er, P\'eter~Makk, C\u at\u alin~Pa\c scu~Moca, Pascal~Simon, Attila~Virosztek and Gergely~Zar\'and for useful discussions. We also acknowledge SNI NanoImaging Lab for FIB cutting, and  M.~G.~Beckern\'e, F.~F\"ul\"op, and M.~Hajd\'u for their technical support.

This research was supported by the Ministry of Innovation and Technology and the National Research, Development and Innovation
Office within the Quantum Information National Laboratory of Hungary and by the Quantum Technology National Excellence Program (Project Nr. 2017-1.2.1-NKP-2017-00001), by the NKFIH fund TKP2020 IES (Grant No.~BME-IE-NAT), and by the OTKA Grant No.~132146, by QuantERA SuperTop project 127900, by AndQC FetOpen project, by SuperGate FetOpen project, by Nanocohybri COST Action CA16218, and by the Danish National Research Foundation. C.~S. acknowledges support from the Swiss National Science Foundation through grants 192027, the NCCR QSIT, NCCR SPIN and the QuantERA project SuperTop. C.S. further acknowledges support from the  European Union’s Horizon 2020 research and innovation programme through grant agreement No 828948, project AndQC. G.~F. acknowledges Bolyai J\'anos Scholarship and was supported by the \'UNKP-20-5 New National Excellence Program of the Ministry for Innovation and Technology from the source of the National Research, Development and Innovation Fund.

\bibliography{scheriff_Shiba2}

\end{document}